\documentclass[12pt,epsf,fleqn]{article}
\usepackage{epsfig}
\usepackage[dvips]{color}
\begin{document}
\title{Higgs bosons of a supersymmetric $E_6$ model at the Large Hadron Collider}
\author{S. W. Ham$^{(1)}$\footnote{swham@knu.ac.kr}, J. O. Im$^{(2)}$\footnote{todd1313@konkuk.ac.kr},
E. J. Yoo$^{(2)}$\footnote{ejyoo01@hanmail.net}, and S. K. Oh$^{(2,3)}$\footnote{sunkun@konkuk.ac.kr}
\\
\\
{\it $^{(1)}$ Department of Physics, KAIST, Daejeon 305-701, Korea} \\
{\it $^{(2)}$ Department of Physics, Konkuk University, Seoul 143-701, Korea} \\
{\it $^{(3)}$ Center for High Energy Physics, Kyungpook National University}  \\
{\it Daegu 702-701, Korea}  \\
\\
\\
}
\date{}
\maketitle
\begin{abstract}
It is found that CP symmetry may be explicitly broken in the Higgs sector of a supersymmetric $E_6$ model
with two extra neutral gauge bosons at the one-loop level.
The phenomenology of the model, the Higgs sector in particular,
is studied for a reasonable parameter space of the model,
in the presence of explicit CP violation at the one-loop level.
At least one of the neutral Higgs bosons of the model might be produced via the $WW$ fusion process
at the Large Hadron Collider.   
\end{abstract}
\vfil\eject

\section{Introduction}

Anticipations among high energy physicists for the discovery of new physics
at the Large Hadron Collider (LHC) are very high as it prepares to operate in full swing.
There are a number of compelling rationales for anticipating new physics beyond the Standard Model (SM).
One of them is the observed baryon asymmetry of the universe,
which indicates the survival of more matter than antimater during the evolution of the universe.
In the SM, the only source of CP violation is the complex phase
in the Cabibbo-Kobayashi-Maskawa (CKM) matrix.
It has been noticed that the size of CP violation in the SM by means of the CKM matrix alone is
too small to explain the observed value of the baryon-to-entropy ratio,
$n_B/s \sim 8 \times 10^{-11}$ [1], if the universe had begun from a baryon-symmetric state.
Thus, in order to explain the observed baryon asymmetry of the universe,
other sources of adequate CP violation are required.

A number of alternative models beyond the SM have been investigated for the possibility of CP violations.
Supersymmetry (SUSY) has been with us for several decades,
which nowadays is regarded as the most certain candidate for new physics.
In reality, the necessity of CP violation beyond the SM is not the only {\it raison d'etre} for the SUSY.
There are multiples of arguments that support its existence in nature.
Some supersymmetric models have also been studied in this context,
as their sophisticated Higgs sectors may possess sources of CP violation [2].
For some phenomenologically realistic supersymmetric models extended from the SM,
soft SUSY breaking terms are essential ingredients [3].
If these soft SUSY breaking terms contain complex phases,
the phenomenological analyses of these supersymmetric standard models might
not only be complicated but also involve CP violation.

The minimal supersymmetric standard model (MSSM) is
the simplest version of supersymmetric extension of the SM.
Its Higgs sector has two Higgs doublets in order to give masses
to up-like quarks and down-like quarks separately.
At the one-loop level, a complex phase in the soft SUSY breaking terms of the MSSM induces
an explicit CP mixing between scalar and pseudoscalar Higgs bosons [4].

Non-minimal versions of supersymmetric extension of the SM have additional Higgs singlets
and thus can dynamically solve the dimensional $\mu$-parameter problem
in the MSSM by means of the vacuum expectation value (VEV) of the Higgs singlet [5,6,7].
They have also been studied within the context of explicit CP violation
in their Higgs sectors [8,9,10,11].
The next-to-minimal supersymmetric standard model (NMSSM) is a typical member of them.
Unlike the MSSM, the Higgs potential of the NMSSM has one nontrivial CP phase
after redefining the Higgs fields at the tree level [9].
At the one-loop level, it also develops CP violating phases.
The effects of explicit CP violation at the one-loop level
in the NMSSM on the masses of neutral and charged Higgs bosons are predicted
in the literature [10].

The Higgs potentials of both the minimal non-minimal supersymmetric model
and the $U(1)$-extended supersymmetric model may not have any CP phase
at the tree level [11].
However, these models may also possess complex phases to induce explicit CP violation
at the one-loop level, by taking the radiative corrections
due to the quark and squark loops into account.

In this article, we would like to continue to study the possibility of CP violation
in the Higgs sector of a supersymmetric $E_6$ model.
This model has two $U(1)$ symmetries in addition to the SM gauge symmetry,
thus with two additional neutral gauge bosons, and two Higgs singlets
as well as two Higgs doublets [12,13].
The tree-level Higgs potential of this model may not have complex phase,
because any complex phase can always be eliminated by rotating the relevant Higgs fields.
At the one-loop level, it is shown that this model may allow CP violation
in an explicit way due to radiative corrections.
We study the Higgs phenomenology of this model by varying all the relevant parameters
within reasonable ranges, to obtain the upper bound on the lightest neutral Higgs boson mass.
We investigate prospects for discovering the neutral Higgs bosons of this model at the LHC,
by calculating the minimum cross section for producing
at least any one of the neutral Higgs bosons of this model via the $WW$ fusion process at the LHC.

\section{Higgs Sector}

Let us describe the Higgs sector of our model.
We assume that the electroweak gauge symmetry of our model is
$SU(2) \times U(1) \times U_1 (1) \times U_2 (1)$,
where the two extra $U(1)$ symmetries are decomposed from $E_6$.
Thus, it is a kind of rank-6 supersymmetric model.
We assume that in general $U_1 (1)$ and $U_2 (1)$ would mix
with a certain mixing angle $\theta$ to become two linearly orthogonal combinations, $U(1)'$ and $U(1)''$.
The Higgs sector of our model consists of two Higgs doublets, $H^T_1 = (H_1^0, H_1^-)$
and $H^T_2 = (H_2^+, H_2^0)$, and two neutral Higgs singlets, $N_1$ and $N_2$.
The Yukawa interaction between Higgs superfields and quark superfields
in the superpotential of our model may be expressed as [12,13]
\begin{equation}
{\cal W} \approx h_t Q^T {\cal H}_2 t_R^c - h_b Q^T {\cal H}_1 b_R^c
+ \lambda {\cal H}_1 {\cal H}_2 {\cal N}_1  \ ,
\end{equation}
where we take only the third generation into account and $h_t$ and $h_b$ are
respectively the dimensionless Yukawa coupling coefficients of top and bottom quarks,
$\lambda$ is a dimensionless coefficient,
${\cal H}_1$ and ${\cal H}_2$ are the Higgs doublet superfields,
${\cal N}$ is the Higgs singlet superfield,
$t_R^c$ and $b_R^c$ are respectively the right-handed top and bottom quark superfields, and
$Q$ is the left-handed $SU(2)$ doublet quark superfield of the third generation.
This superpotential has the same expression as discussed in Ref. [13] or Ref. [14],
where relatively well-known rank-6 SUSY models are investigated.

Note that, shown as the last term in the superpotential,
only $N_1$ participates in coupling to the Higgs doublets,
because the underlying $E_6$ gauge symmetry does not allow the other Higgs singlet $N_2$ to do so [13].
Effectively, the coupling between $N_1$ and the Higgs doublets corresponds to the $\mu$ term in the MSSM
where the $\mu$-parameter is generated by the VEV of the $N_1$.

The Higgs potential of our model at the tree level is collected from $D$-terms, $F$-terms,
and the soft terms in the superpotential.
The most general form of the Higgs potential at the tree level is given as [13]
\begin{eqnarray}
V_0 & = & m_1^2 H_1^{\dagger} H_1 + m_2^2 H_2^{\dagger} H_2 + m_3^2 N_1^{\dagger} N_1
+ m_4^2 N_2^{\dagger} N_2 - (\lambda A H_1 H_2 N_1 + {\rm H.c.} )  \cr
&  &\mbox{} + |\lambda|^2 [H_1^{\dagger} H_1 H_2^{\dagger} H_2 + H_1^{\dagger} H_1 N_1^{\dagger} N_1
+ H_2^{\dagger} H_2 N_1^{\dagger} N_1 ]  \  \cr
&  &\mbox{} + \left ( {g_2^2 \over 2} - |\lambda|^2 \right )  |H_1^{\dagger} H_2|^2
 + {g_1^2 + g_2^2 \over 8} (H_1^{\dagger} H_1 - H_2^{\dagger} H_2)^2  \cr
& &\mbox{} + {g_1^{'2} \over 72} [ C_{\theta} (H_1^{\dagger} H_1 + 4 H_2^{\dagger} H_2
- 5 N_1^{\dagger} N_1 - 5 N_2^{\dagger} N_2 ) \cr
& &\mbox{} - \sqrt{15} S_{\theta} (H_1^{\dagger} H_1
- N_1^{\dagger} N_1 + N_2^{\dagger} N_2  ) ]^2  \cr
& &\mbox{} + {g_1^{''2} \over 72} [ S_{\theta} (H_1^{\dagger} H_1 + 4 H_2^{\dagger} H_2
- 5 N_1^{\dagger} N_1 - 5 N_2^{\dagger} N_2 ) \cr
& & \mbox{} + \sqrt{15} C_{\theta} (H_1^{\dagger} H_1
- N_1^{\dagger} N_1 + N_2^{\dagger} N_2  ) ]^2   \   ,
\end{eqnarray}
where $g_2$, $g_1$, $g_1^{'}$, and $g_1^{''}$ are
respectively the $SU(2)$, $U(1)$, $U(1)'$, and $U(1)''$ gauge coupling coefficients,
$A$ is a massive parameter, $C_{\theta} = \cos \theta$ and $S_{\theta} = \sin \theta$,
and $m_i$ ($i=1,2,3,4$) are soft SUSY breaking masses.
These four soft masses in the Higgs potential would eventually be eliminated
by means of the minimum conditions for the Higgs potential with respect to four neutral Higgs fields.

The parameters of the Higgs potential are assumed to be generally complex.
Thus, $\lambda$ and  $A$ in the tree-level Higgs potential may be complex
such that their complex phases may be factored out explicitly as $\lambda A e^{i\phi}$.
We also assume that the VEVs, which four neutral components of the Higgs fields acquire
after electroweak symmetry breaking, may in general be complex.
However, by redefining the phases of $H_1$, $H_2$, and $N_2$,
we may adjust the vacuum expectation values as $v_1 = \langle H_1^0 \rangle$,
$v_2 = \langle H_2^0 \rangle$, $x_1 e^{i \phi_1} = \langle N_1 \rangle$
and $x_2 = \langle N_2 \rangle$, where $v_1$, $v_2$, $x_1$, and $x_2$ are real
and the complex phase $\phi_1$ is the overall phase in $\langle H_1 H_2 N_1 \rangle$.
Thus, looking at the Higgs potential at the tree level, one can easily notice
that the only possible source of complex phases is $\lambda A H_1H_2N$.
By further redefining the phase of the Higgs singlet $N_1$,
it is always possible to make the two phases $\phi$ and $\phi_1$ cancel
each other so that the tree-level Higgs potential can be made completely real.
Therefore, our model conserves the CP symmetry at the tree level.

After the electroweak symmetry breaking, the tree-level mass of top quark is
given as $m_t^2 = (h_t v_2)^2$, and the tree-level masses of stop quarks are
given by the on-shell Lagrangian as
\begin{equation}
m_{{\tilde t}_1, \ {\tilde t}_2}^2 =  {1 \over 2} (m_Q^2 + m_T^2) + m_t^2
+ {1 \over 4} m_Z^2 \cos 2 \beta + G_t^{'}
\mp \sqrt{X_t} \ ,
\end{equation}
where $m_Q$ and $m_T$ are the soft SUSY breaking masses for the stop quarks,
$m^2_Z = (g^2_1+g^2_2)v^2 /2$ with $v^2 = v^2_1+v^2_2$ is
the squared mass of the neutral weak gauge boson, $\tan\beta = v_2/v_1$, and
\begin{eqnarray}
X_t & = & \left ( {1 \over 2} (m_Q^2 - m_T^2) + \left ({2 \over 3} m_W^2 - {5 \over 12} m_Z^2 \right ) \cos 2 \beta \right)^2 \cr
& &\mbox{} + m_t^2 ( A_t^2 + \lambda^2 x_1^2 \cot^2 \beta - 2 \lambda A_t x_1 \cot \beta \cos \phi_t ) \ , \cr
G_t^{'} & = &\mbox{} - {g_1^{' 2} \over 4} \left ( {1 \over 3} \sqrt{5 \over 2} S_{\theta} - {1 \over \sqrt{6} } C_{\theta} \right )
\left [ \left ( {\sqrt{10} \over 3} S_{\theta} + \sqrt{2 \over 3} C_{\theta} \right ) v^2 \cos^2 \beta
- {2 \over 3} \sqrt{10} S_{\theta} x_1^2  \right. \cr
& &\left. \mbox{} + \left ( {\sqrt{10} \over 3} S_{\theta} - \sqrt{2 \over 3} C_{\theta} \right ) v^2 \sin^2 \beta
- \left ( {1 \over 3} \sqrt{5 \over 2} S_{\theta} - {5 \over \sqrt{6}} C_{\theta} \right ) x_2^2 \right ]  \cr
& &\mbox{} - {g_1^{'' 2} \over 4} \left ( {1 \over 3} \sqrt{5 \over 2} C_{\theta} + {1 \over \sqrt{6} } S_{\theta} \right )
\left [ \left ( {\sqrt{10} \over 3} C_{\theta} - \sqrt{2 \over 3} S_{\theta} \right ) v^2 \cos^2 \beta
- {2 \over 3} \sqrt{10} C_{\theta} x_1^2  \right. \cr
& &\left. \mbox{} + \left ( {\sqrt{10} \over 3} C_{\theta} + \sqrt{2 \over 3} S_{\theta} \right ) v^2 \sin^2 \beta
- \left ( {1 \over 3} \sqrt{5 \over 2} C_{\theta} + {5 \over \sqrt{6}} S_{\theta} \right ) x_2^2 \right ]    \  ,
\end{eqnarray}
with $m^2_W = g^2_2v^2/2$ being the squared mass of the charged weak gauge boson,
$A_t$ being the trilinear soft SUSY breaking parameter of the stop quarks
with mass dimension, and $\phi_t$ being a complex phase determined by $\phi_1$
and the complex phase of $A_t$.
Note that $G_t^{'}$ is the effect of the two extra $U(1)$ symmetries,
but it does not contribute the mass splitting between the two stop quark masses.
The mixing, and hence the mass splitting, between the stop quark masses is triggered by $X_t$.

Now let us consider the one-loop radiative corrections to the tree-level Higgs potential.
In supersymmetric models, the incomplete cancellation between ordinary particles
and their superpartners yield the one-loop corrections to the tree-level Higgs boson masses.
In SUSY models, the most dominant part of the one-loop corrections to the tree-level Higgs potential
come primarily from the top and stop quark loops.
For large $\tan \beta$ as large as 50, the contribution of the bottom and sbottom quark loops
can also be large.
In this paper, we consider the contributions from the top and stop quark loops
at the one-loop level.
The full Higgs potential at the one-loop level may be written as $V = V_0 + V_1$,
where $V_1$ is contribution from the radiative corrections due to the top and stop quark loops.
The effective potential method provides us [15]
\begin{equation}
    V_1 = \sum_{l} {n_l {\cal M}_l^4 \over 64 \pi^2}
    \left [ \log {{\cal M}_l^2 \over \Lambda^2} - {3 \over 2} \right ]  \ ,
\end{equation}
where $\Lambda$ is the renormalization scale in the modified minimal subtraction scheme,
the subscript $l$ stands for the top and stop quarks: $t$, ${\tilde t}_1$, ${\tilde t}_2$,
${\cal M}_i$ are the top and stop masses at the tree level given as functions of Higgs fields,
and $n_i$ are the degrees of freedom for these particles.
Including the sign convention, they are given as $n_t = -12$ and
$n_{{\tilde t}_i} = 6$ ($i=1,2$),
as in the above formula enter the stop quarks with a negative sign while the top quark with a positive sign.

Since the parameters of the Higgs potential are assumed to be generally complex, we may have
$\phi$, which is the phase of $\lambda A$.
Unlike the tree-level case, we cannot redefine the phase of $N_1$ at the one-loop level to cancel it.
Thus, $\phi$ may persist at the one-loop level.
This can be most clearly be seen in the  non-trivial tadpole minimum condition
with respect to the pseudoscalar component of the Higgs field:
\begin{equation}
0 = A \sin \phi
- {3 m_t^2 A_t \sin \phi_t \over 16 \pi^2 v^2 \sin^2 \beta} f (m_{{\tilde t}_1}^2,  \ m_{{\tilde t}_2}^2)   \ ,
\end{equation}
where the first term comes from the tree-level Higgs potential
and the second term comes from the radiative corrections,
and the dimensionless function $f$ arising from radiative corrections is defined as
\begin{equation}
 f(m_x^2, \ m_y^2) = {1 \over (m_y^2 - m_x^2)} \left[  m_x^2 \log {m_x^2 \over \Lambda^2} - m_y^2
\log {m_y^2 \over \Lambda^2} \right] + 1 \ .
\end{equation}
But for the radiative corrections, the above tadpole minimum condition
at the tree level would be satisfied when $\phi = 0$.
Due to the presence of the one-loop corrections, $\phi=0$ is no longer
in general the solution to the above tadpole minimum condition.

Our model has twelve real degrees of freedom in the Higgs sector.
They may be classified as three neutral Goldstone bosons, a pair of charged Goldstone bosons,
five neutral Higgs bosons and a pair of charged Higgs bosons.
After the electroweak symmetry breaking, the three neutral Goldstone bosons and
a pair of charged Goldstone bosons will be absorbed
into the longitudinal component of $Z$, $Z'$, $Z''$ and $W$ gauge bosons,
where $Z'$ and $Z''$ are the extra neutral gauge bosons.

The squared mass matrix $M$ of the five neutral Higgs bosons is given
as a symmetric $5 \times 5$ matrix,
obtained by the second derivatives of the Higgs potential
with respect to the five neutral Higgs fields.
At the tree level, the five neutral Higgs bosons may have definite CP parity,
since the CP symmetry is conserved in the Higgs sector.
Thus, we may denote them as $S_i$ ($i =1,2,3,4$)
for neutral scalar Higgs bosons and $P$ for neutral pseudoscalar Higgs boson.
In the ($S_1, S_2, S_3, S_4, P$) basis, the $5 \times 5$ matrix $M$
at the tree level may be expressed as
\begin{equation}
	M = M^0 + M^{0'}    \   ,
\end{equation}
where $M^{0'}$ comes from the $D$-terms due to two extra $U(1)$ symmetries of $V^0$, and
$M^0$ comes from the remaining terms in $V_0$,
namely, $D$-terms due to the SM gauge symmetry, the $F$-terms,
and the soft terms of the tree-level Higgs potential.
They may be expressed as
\begin{equation}
     M^0 =
    \left ( \begin{array}{ccccc}
    M_{11}^0 & M_{12}^0 & M_{13}^0 & 0 & 0 \cr
    M_{12}^0 & M_{22}^0 & M_{23}^0 & 0 & 0  \cr
    M_{13}^0 & M_{23}^0 & M_{33}^0 & 0 & 0  \cr
    0        & 0        & 0        & 0 & 0  \cr
    0        & 0        & 0        & 0 & M_{55}^0
        \end{array}
    \right ) \ ,
\end{equation}
\begin{equation}
     M^{0'} =
    \left ( \begin{array}{ccccc}
    M_{11}^{0'} & M_{12}^{0'} & M_{13}^{0'} & M_{14}^{0'} & 0 \cr
    M_{12}^{0'} & M_{22}^{0'} & M_{23}^{0'} & M_{24}^{0'} & 0  \cr
    M_{13}^{0'} & M_{23}^{0'} & M_{33}^{0'} & M_{34}^{0'} & 0  \cr
    M_{14}^{0'} & M_{24}^{0'} & M_{34}^{0'} & M_{44}^{0'} & 0  \cr
    0           & 0           &        0    & 0           &  0
        \end{array}
    \right ) \ ,
\end{equation}
Explicitly, the matrix elements of $M^0$ and $M^{0'}$ are respectively given as follows:
\begin{eqnarray}
M_{11}^0 & = & m_Z^2 \cos^2 \beta + M_{55}^0 \sin^2 \beta \cos^2 \alpha  \ ,  \cr
M_{22}^0 & = & m_Z^2 \sin^2 \beta + M_{55}^0 \cos^2 \beta \cos^2 \alpha \ ,  \cr
M_{33}^0 & = & M_{55}^0 \sin^2 \alpha \ , \cr
M_{12}^0 & = & (\lambda^2 v^2 - m_Z^2/2) \sin 2 \beta - M_{55}^0 \cos \beta \sin \beta \cos^2 \alpha \ ,  \cr
M_{13}^0 & = & 2 \lambda^2 v x_1 \cos \beta - M_{55}^0 \sin \beta \cos \alpha \sin \alpha \ , \cr
M_{23}^0 & = & 2 \lambda^2 v x_1 \sin \beta - M_{55}^0 \cos \beta \cos \alpha \sin \alpha  \ , \cr
M_{55}^0 & = & 2 \lambda A v {\cos \phi  \over \sin 2 \alpha}  \  ,
\end{eqnarray}
and
\begin{eqnarray}
M_{11}^{0'} & = & {1 \over 18} (g_1^{'2} C_{\theta}^2 + g_1^{''2}S_{\theta}^2 ) v^2 \cos^2 \beta
+ {5 \over 6} (g_1^{'2} S_{\theta}^2 + g_1^{''2}C_{\theta}^2) v^2 \cos^2 \beta \cr
& &\mbox{} - {\sqrt{15} \over 9} (g_1^{'2} - g_1^{''2}) C_{\theta} S_{\theta} v^2 \cos^2 \beta   \ ,  \cr
M_{22}^{0'} & = &  {8 \over 9} (g_1^{'2} C_{\theta}^2 + g_1^{''2}S_{\theta}^2 ) v^2 \sin^2 \beta \ ,  \cr
M_{33}^{0'} & = &  {25 \over 18} (g_1^{'2} C_{\theta}^2 + g_1^{''2}S_{\theta}^2 ) x_1^2
+ {5 \over 6} (g_1^{'2} S_{\theta}^2 + g_1^{''2}C_{\theta}^2) x_1^2
- {5 \sqrt{15} \over 9} (g_1^{'2} - g_1^{''2}) C_{\theta} S_{\theta} x_1^2   \ ,  \cr
M_{44}^{0'} & = &  {25 \over 18} (g_1^{'2} C_{\theta}^2 + g_1^{''2}S_{\theta}^2 ) x_2^2
+ {5 \over 6} (g_1^{'2} S_{\theta}^2 + g_1^{''2}C_{\theta}^2) x_2^2
+ {5 \sqrt{15} \over 9} (g_1^{'2} - g_1^{''2}) C_{\theta} S_{\theta} x_2^2   \ ,  \cr
M_{12}^{0'} & = & {1 \over 9} (g_1^{'2} C_{\theta}^2 + g_1^{''2}S_{\theta}^2 ) v^2 \sin 2 \beta
- {\sqrt{15} \over 9} (g_1^{'2} - g_1^{''2}) C_{\theta} S_{\theta} v^2 \sin 2 \beta    \ ,  \cr
M_{13}^{0'} & = &\mbox{} - {5 \over 18} (g_1^{'2} C_{\theta}^2 + g_1^{''2}S_{\theta}^2 ) v x_1 \cos \beta
- {5 \over 6} (g_1^{'2} S_{\theta}^2 + g_1^{''2}C_{\theta}^2) v x_1 \cos \beta \cr
& &\mbox{} + {\sqrt{15} \over 3} (g_1^{'2} - g_1^{''2}) C_{\theta} S_{\theta} v x_1 \cos \beta    \ ,  \cr
M_{14}^{0'} & = &\mbox{} - {5 \over 18} (g_1^{'2} C_{\theta}^2 + g_1^{''2}S_{\theta}^2 ) v x_2 \cos \beta
+ {5 \over 6} (g_1^{'2} S_{\theta}^2 + g_1^{''2}C_{\theta}^2) v x_2 \cos \beta \cr
& &\mbox{} + {2 \sqrt{15} \over 3} (g_1^{'2} - g_1^{''2}) C_{\theta} S_{\theta} v x_2 \cos \beta    \ ,  \cr
M_{23}^{0'} & = &\mbox{} - {10 \over 9} (g_1^{'2} C_{\theta}^2 + g_1^{''2}S_{\theta}^2 ) v x_1 \sin \beta
+ {2 \sqrt{15} \over 9} (g_1^{'2} - g_1^{''2}) C_{\theta} S_{\theta} v x_1 \sin \beta     \ ,  \cr
M_{24}^{0'} & = &\mbox{} - {10 \over 9} (g_1^{'2} C_{\theta}^2 + g_1^{''2}S_{\theta}^2 ) v x_2 \sin \beta
- {2 \sqrt{15} \over 9} (g_1^{'2} - g_1^{''2}) C_{\theta} S_{\theta} v x_2 \sin \beta       \ ,  \cr
M_{34}^{0'} & = & {25 \over 18} (g_1^{'2} C_{\theta}^2 + g_1^{''2}S_{\theta}^2 ) x_1 x_2
- {5 \over 6} (g_1^{'2} S_{\theta}^2 + g_1^{''2}C_{\theta}^2) x_1 x_2 \ ,
\end{eqnarray}
where $\tan \alpha = (v / 2 x_1) \sin 2 \beta$ stands for the splitting
between an extra $U(1)$ symmetry breaking scale and the electroweak scale.

Note that both $M^0$ and $M^{0 '}$ do not mix $S_i$ with $P$.
In other words, there is no scalar-psuedoscalar mixing at the tree-level,
hence the CP symmetry.
It is straightforward to recognize that the single element $M^0_{55}$ is the squared mass
at the tree level of the neutral pseudoscalar Higgs boson.
Note also that if the two extra $U(1)$ symmetries are absent, we would have $M^{0'} = 0$.
In this case, one of the neutral scalar Higgs bosons would be massless at the tree level,
since $M^0$ may be decomposed into a block diagram consisting of three blocks,
namely, $3\times3$ submatrix, $M^0_{44} = 0$ and $M^0_{55}$.

Now, at the one-loop level, the squared mass matrix $M$ of the five neutral Higgs bosons is corrected as
\begin{equation}
M = M^0 + M^{0'} + M^1   \   ,
\end{equation}
where $M^1$ is the radiative corrections obtained from $V^1$ as
\begin{equation}
     M^1 =
    \left ( \begin{array}{ccccc}
    M_{11}^1 & M_{12}^1 & M_{13}^1  & M_{14}^1  & M_{15}^1   \cr
    M_{12}^1 & M_{22}^1 & M_{23}^1  & M_{24}^1  & M_{25}^1   \cr
    M_{13}^1 & M_{23}^1 & M_{33}^1  & M_{34}^1  & M_{35}^1   \cr
    M_{14}^1 & M_{24}^1 & M_{34}^1  & M_{44}^1  & M_{45}^1   \cr
    M_{15}^1 & M_{25}^1 & M_{35}^1  & M_{45}^1  & M_{55}^1
       \end{array}
          \right ) \ .
\end{equation}
Explicitly, the matrix elements of $M^1$ are given as follows, after imposing tadpole minimum conditions:
\begin{eqnarray}
M_{11}^1 & = & m_A^2 \sin^2 \beta \cos^2 \alpha  - {3 \cos^2 \beta \over 16 \pi^2 v^2}
\left( {4 m_W^2 \over 3} - {5 m_Z^2 \over 6} \right)^2
f(m_{{\tilde t}_1}^2, \ m_{{\tilde t}_2}^2) \cr
& & \cr
& &\mbox{} + {3 \over 8 \pi^2 v^2}
\left ( {m_t^2 \lambda x_1 \Delta_{{\tilde t}_1} \over \sin \beta} + {\cos \beta \Delta_{\tilde t} \over 2} \right )^2
{g(m_{{\tilde t}_1}^2, \ m_{{\tilde t}_2}^2) \over (m_{{\tilde t}_2}^2 - m_{{\tilde t}_1}^2)^2} \cr
& & \cr
& &\mbox{} + {3 \cos^2 \beta \over 128 \pi^2 v^2} ( 4 G_a v^2 + m_Z^2 )^2
\log \left ({m_{{\tilde t}_1}^2  m_{{\tilde t}_2}^2 \over \Lambda^4} \right ) \cr
& & \cr
& &\mbox{}
+ {3 \cos \beta \over 16 \pi^2 v^2} (4 G_a v^2 + m_Z^2)
\left ( {m_t^2 \lambda x_1 \Delta_{{\tilde t}_1} \over \sin \beta} + {\cos \beta \Delta_{\tilde t} \over 2} \right )
{\displaystyle \log (m_{{\tilde t}_2}^2 / m_{{\tilde t}_1}^2)  \over (m_{{\tilde t}_2}^2 - m_{{\tilde t}_1}^2)}
 \ , \cr
& & \cr
M_{22}^1 & = & m_A^2 \cos^2 \beta \cos^2 \alpha
- {3 \sin^2 \beta \over 16 \pi^2 v^2}
\left( {4 m_W^2 \over 3} - {5 m_Z^2 \over 6} \right)^2
f(m_{{\tilde t}_1}^2, \ m_{{\tilde t}_2}^2)     \cr
& & \cr
& &\mbox{} + {3 \sin^2 \beta \over 8 \pi^2 v^2}
\left ( {m_t^2 A_t \Delta_{{\tilde t}_2} \over \sin^2 \beta} + {\Delta_{\tilde t} \over 2} \right )^2
{g(m_{{\tilde t}_1}^2, \ m_{{\tilde t}_2}^2) \over(m_{{\tilde t}_2}^2 - m_{{\tilde t}_1}^2)^2}
- {3 m_t^4 \over 4 \pi^2 v^2 \sin^2 \beta} \log \left ({m_t^2 \over \Lambda^2} \right )  \cr
& & \cr
& &\mbox{} - {3 \sin^2 \beta \over 16 \pi^2 v^2}
\left ({4 m_t^2 \over \sin^2 \beta} - m_Z^2 + 4 G_b v^2 \right)
\left ( {m_t^2 A_t \Delta_{{\tilde t}_2} \over \sin^2 \beta}
+ {\Delta_{\tilde t} \over 2} \right )
{\displaystyle \log (m_{{\tilde t}_2}^2 / m_{{\tilde t}_1}^2)
 \over (m_{{\tilde t}_2}^2 - m_{{\tilde t}_1}^2)} \cr
& & \cr
& &\mbox{}
 + {3 \sin^2 \beta \over 32 \pi^2 v^2}
\left ({2 m_t^2 \over \sin^2 \beta} - {m_Z^2 \over 2} + 2 G_b v^2 \right)^2
\log \left ({m_{{\tilde t}_1}^2  m_{{\tilde t}_2}^2 \over \Lambda^4} \right ) \ , \cr
& & \cr
M_{33}^1 & = &   m_A^2 \sin^2 \alpha + {3 m_t^4 \lambda^2 {\Delta_{{\tilde t}_1}^2}
\over 8 \pi^2 \tan^2 \beta}
{g(m_{{\tilde t}_1}^2, \ m_{{\tilde t}_2}^2) \over (m_{{\tilde t}_2}^2 - m_{{\tilde t}_1}^2)^2 }
- {3 G_c m_t^2 x_1 \lambda \Delta_{{\tilde t}_1}  \over 4 \pi^2 \tan \beta}
{\displaystyle \log (m_{{\tilde t}_2}^2 / m_{{\tilde t}_1}^2)
 \over (m_{{\tilde t}_2}^2 - m_{{\tilde t}_1}^2)} \cr
& & \cr
& &\mbox{}
 + {3 G_c^2 x_1^2 \over 8 \pi^2}
\log \left ({m_{{\tilde t}_1}^2  m_{{\tilde t}_2}^2 \over \Lambda^4} \right ) \ , \cr
& & \cr
M_{44}^1 & = & {3 G_d^2 x_2^2 \over 8 \pi^2}
\log \left ({m_{{\tilde t}_1}^2  m_{{\tilde t}_2}^2 \over \Lambda^4} \right ) \ , \cr
& & \cr
M_{55}^1 & = &  m^2_A + {3 m_t^4 \lambda^2 A_t^2 x_1^2 \sin^2 \phi_t \over 8 \pi^2 v^2 \sin^4 \beta \cos^2 \alpha}
{g(m_{{\tilde t}_1}^2, \ m_{{\tilde t}_2}^2) \over (m_{{\tilde t}_2}^2 - m_{{\tilde t}_1}^2 )^2}  \ , \cr
& & \cr
M_{12}^1 & = &\mbox{} - m_A^2 \cos \beta \sin \beta \cos^2 \alpha
 + {3 \sin 2 \beta \over 32 \pi^2 v^2} \left ({4 m_W^2 \over 3} - {5 m_Z^2 \over 6} \right)^2
f(m_{{\tilde t}_1}^2, \ m_{{\tilde t}_2}^2) \cr
& & \cr
& &\mbox{} - {3 \sin \beta \over 8 \pi^2 v^2}
\left ( {m_t^2 \lambda x_1 \Delta_{{\tilde t}_1} \over \sin \beta} + {\cos \beta \Delta_{\tilde t} \over 2} \right )
\left ( {m_t^2 A_t \Delta_{{\tilde t}_2}\over \sin^2 \beta} + {\Delta_{\tilde t} \over 2} \right )
{g(m_{{\tilde t}_1}^2, \ m_{{\tilde t}_2}^2)
\over (m_{{\tilde t}_2}^2 - m_{{\tilde t}_1}^2)^2} \cr
& & \cr
& &\mbox{} + {3 \sin 2 \beta \over 32 \pi^2 v^2}
\left ({4 m_t^2 \over \sin^2 \beta} - m_Z^2 + 4 G_b v^2 \right)
\left ({m_t^2 \lambda x_1 \Delta_{{\tilde t}_1} \over \sin 2\beta}
+ {\Delta_{\tilde t} \over 4} \right )
{\displaystyle \log (m_{{\tilde t}_2}^2 / m_{{\tilde t}_1}^2) \over (m_{{\tilde t}_2}^2 - m_{{\tilde t}_1}^2)} \cr
& & \cr
& &\mbox{} - {3 \sin 2 \beta \over 64 \pi^2 v^2} (4 G_a v^2 + m_Z^2)
\left ({m_t^2 A_t \Delta_{{\tilde t}_2} \over \sin^2 \beta}
+ {\Delta_{\tilde t} \over 2} \right )
{\displaystyle \log (m_{{\tilde t}_2}^2 / m_{{\tilde t}_1}^2)
 \over (m_{{\tilde t}_2}^2 - m_{{\tilde t}_1}^2)} \cr
& & \cr
& &\mbox{}
+ {3 \sin 2 \beta \over 256 \pi^2 v^2} (4 G_a v^2 + m_Z^2)
\left ({4 m_t^2 \over \sin^2 \beta} - m_Z^2 + 4 G_b v^2 \right)
\log \left ({m_{{\tilde t}_1}^2 m_{{\tilde t}_2}^2 \over \Lambda^4} \right ) \ , \cr
& & \cr
M_{13}^1 & = &\mbox{} - m_A^2 \sin \beta \cos \alpha \sin \alpha
- {3 m_t^2 \lambda^2 x_1 \cos \beta \over 8 \pi^2 v \sin^2 \beta}
f(m_{{\tilde t}_1}^2, \ m_{{\tilde t}_2}^2)  \cr
& &  \cr
& &\mbox{}  +  {3 m_t^2 \lambda \Delta_{{\tilde t}_1}\over 8 \pi^2 v \tan \beta}
\left ( {m_t^2 \lambda x_1 \Delta_{{\tilde t}_1} \over \sin \beta}
+ {\cos \beta \Delta_{\tilde t} \over 2} \right )
{g(m_{{\tilde t}_1}^2, \ m_{{\tilde t}_2}^2) \over (m_{{\tilde t}_2}^2 - m_{{\tilde t}_1}^2)^2 }   \cr
& & \cr
& &\mbox{} + {3 m_t^2 \lambda \cos \beta\Delta_{{\tilde t}_1} \over 32 \pi^2 v \tan \beta}
( 4 G_a v^2 + m_Z^2)  {\log( {m_{{\tilde t}_2}^2 / m_{{\tilde t}_1}^2})
\over (m_{{\tilde t}_2}^2 - m_{{\tilde t}_1}^2)}  \cr
& & \cr
& &\mbox{} + {3 G_c x_1 \over 8 \pi^2 v}
\left ({m_t^2 \lambda x_1 \Delta_{{\tilde t}_1} \over \sin \beta}
+ {\cos \beta \Delta_{\tilde t} \over 2} \right )
{\displaystyle \log (m_{{\tilde t}_2}^2 / m_{{\tilde t}_1}^2)
 \over (m_{{\tilde t}_2}^2 - m_{{\tilde t}_1}^2)} \cr
& & \cr
& &\mbox{}
+ {3 G_c x_1 \cos \beta \over 32 \pi^2 v} (4 G_a v^2 + m_Z^2)
\log \left ({m_{{\tilde t}_1}^2 m_{{\tilde t}_2}^2 \over \Lambda^4} \right ) \ , \cr
& & \cr
M_{14}^1 & = & \mbox{} {3 G_d x_2 \over 8 \pi^2 v}
\left ({m_t^2 \lambda x_1 \Delta_{{\tilde t}_1} \over \sin \beta}
+ {\cos \beta \Delta_{\tilde t} \over 2} \right )
{\displaystyle \log (m_{{\tilde t}_2}^2 / m_{{\tilde t}_1}^2)
 \over (m_{{\tilde t}_2}^2 - m_{{\tilde t}_1}^2)} \cr
& & \cr
& &\mbox{}
+ {3 G_d x_2 \cos \beta \over 32 \pi^2 v} (4 G_a v^2 + m_Z^2)
\log \left ({m_{{\tilde t}_1}^2 m_{{\tilde t}_2}^2 \over \Lambda^4} \right ) \ , \cr
& & \cr
M_{15}^1 & = & {3 m_t^4 \lambda^2 A_t x_1^2 \Delta_{{\tilde t}_1} \sin \phi_t
\over 8 \pi^2 v^2 \sin^3 \beta \cos \alpha}
{g(m_{{\tilde t}_1}^2, \ m_{{\tilde t}_2}^2) \over (m_{{\tilde t}_2}^2 - m_{{\tilde t}_1}^2)^2 } \cr
& &\mbox{} + {3 m_t^2 \lambda A_t \cos \beta \Delta_{\tilde t} \sin \phi_t
\over 16 \pi^2 v \tan \beta \sin \alpha}
{g(m_{{\tilde t}_1}^2, \ m_{{\tilde t}_2}^2) \over (m_{{\tilde t}_2}^2 - m_{{\tilde t}_1}^2)^2 }  \cr
& & \cr
& &\mbox{} - {3 m_t^2 \lambda A_t \cos \beta \sin \phi_t \over 32 \pi^2 v \tan \beta \sin \alpha}
(4 G_a v^2 + m_Z^2)
\ {\log ({m_{{\tilde t}_2}^2 / m_{{\tilde t}_1}^2}) \over (m_{{\tilde t}_2}^2 - m_{{\tilde t}_1}^2) } \ , \cr
& & \cr
M_{23}^1 & = &\mbox{} - m_A^2 \cos \beta \cos \alpha \sin \alpha   \cr
& & \cr
& &\mbox{} - {3 m_t^2 \lambda \Delta_{{\tilde t}_1} \over 8 \pi^2 v \tan \beta}
\left( {m_t^2 A_t \Delta_{{\tilde t}_2} \over \sin \beta} + {\sin \beta \Delta_{\tilde t} \over 2} \right )
{g(m_{{\tilde t}_1}^2, \ m_{{\tilde t}_2}^2) \over (m_{{\tilde t}_2}^2 - m_{{\tilde t}_1}^2)^2} \cr
& & \cr
& &\mbox{} + {3 m_t^2 \lambda \cos \beta \Delta_{{\tilde t}_1} \over 16 \pi^2 v}
\left ({2 m_t^2 \over \sin^2 \beta} - {m_Z^2 \over 2} + 2 G_b v^2 \right)
{\log (m_{{\tilde t}_2}^2 / m_{{\tilde t}_1}^2) \over (m_{{\tilde t}_2}^2 - m_{{\tilde t}_1}^2)}  \cr
& & \cr
& &\mbox{} - {3 G_c x_1 \over 8 \pi^2 v}
\left ({m_t^2 A_t \Delta_{{\tilde t}_2} \over \sin \beta}
+ {\sin \beta \Delta_{\tilde t} \over 2} \right )
{\displaystyle \log (m_{{\tilde t}_2}^2 / m_{{\tilde t}_1}^2)
 \over (m_{{\tilde t}_2}^2 - m_{{\tilde t}_1}^2)} \cr
& & \cr
& &\mbox{}
+ {3 G_c x_1 \sin \beta \over 32 \pi^2 v} \left ({4 m_t^2 \over \sin^2 \beta} + 4 G_b v^2 - m_Z^2 \right)
\log \left ({m_{{\tilde t}_1}^2 m_{{\tilde t}_2}^2 \over \Lambda^4} \right ) \ , \cr
& & \cr
M_{24}^1 & = &\mbox{} - {3 G_d x_2 \over 8 \pi^2 v}
\left ({m_t^2 A_t \Delta_{{\tilde t}_2} \over \sin \beta}
+ {\sin \beta \Delta_{\tilde t} \over 2} \right )
{\displaystyle \log (m_{{\tilde t}_2}^2 / m_{{\tilde t}_1}^2)
 \over (m_{{\tilde t}_2}^2 - m_{{\tilde t}_1}^2)} \cr
& & \cr
& &\mbox{}
+ {3 G_d x_2 \sin \beta \over 32 \pi^2 v} \left ({4 m_t^2 \over \sin^2 \beta} + 4 G_b v^2 - m_Z^2 \right)
\log \left ({m_{{\tilde t}_1}^2 m_{{\tilde t}_2}^2 \over \Lambda^4} \right ) \ , \cr
& & \cr
M_{25}^1 & = &  \mbox{} - {3 m_t^4 \lambda A_t^2 x_1 \Delta_{{\tilde t}_2} \sin \phi_t
\over 8 \pi^2 v^2 \sin^3 \beta \cos \alpha}
{g(m_{{\tilde t}_1}^2, \ m_{{\tilde t}_2}^2) \over (m_{{\tilde t}_2}^2 - m_{{\tilde t}_1}^2)^2 }  \cr
& &\mbox{} - {3 m_t^2 \lambda A_t \cos \beta \Delta_{\tilde t} \sin \phi_t
\over 16 \pi^2 v \sin \alpha}
{g(m_{{\tilde t}_1}^2, \ m_{{\tilde t}_2}^2) \over (m_{{\tilde t}_2}^2 - m_{{\tilde t}_1}^2)^2 }   \cr
& & \cr
& &\mbox{} + {3 m_t^2 \lambda A_t \cos \beta \sin \phi_t \over 32 \pi^2 v \sin \alpha}
\left ({4 m_t^2 \over \sin^2 \beta} + 4 G_b v^2 - m_Z^2 \right )
{\log (m_{{\tilde t}_2}^2 / m_{{\tilde t}_1}^2)
\over (m_{{\tilde t}_2}^2 - m_{{\tilde t}_1}^2)} \ , \cr
& & \cr
M_{34}^1 & = & {3 m_t^2 G_d x_2 \lambda \Delta_{{\tilde t}_1} \over 8 \pi^2 \tan \beta}
{\displaystyle \log (m_{{\tilde t}_2}^2 / m_{{\tilde t}_1}^2)
 \over (m_{{\tilde t}_2}^2 - m_{{\tilde t}_1}^2)}
+ {3 G_c G_d x_1 x_2 \over 8 \pi^2}
\log \left ({m_{{\tilde t}_1}^2 m_{{\tilde t}_2}^2 \over \Lambda^4} \right ) \ , \cr
& & \cr
M_{35}^1 & = & {3 m_t^4 \lambda^2 A_t x_1 \Delta_{{\tilde t}_1} \sin \phi_t
\over 8 \pi^2 v \sin^2 \beta \tan \beta \cos \alpha}
{g(m_{{\tilde t}_1}^2, \ m_{{\tilde t}_2}^2) \over (m_{{\tilde t}_2}^2 - m_{{\tilde t}_1}^2)^2 }  \cr
& & \cr
& &\mbox{} + {3 m_t^2 G_c A_t \lambda v \cos^2 \beta \sin \phi_t \over 8 \pi^2 \tan \alpha \sin \alpha}
{\displaystyle \log (m_{{\tilde t}_2}^2 / m_{{\tilde t}_1}^2)
 \over (m_{{\tilde t}_2}^2 - m_{{\tilde t}_1}^2)}   \  , \cr
& & \cr
M_{45}^1 & = &  {3 m_t^2 G_d A_t x_2 \lambda \sin \phi_t \over 8 \pi^2 \tan \beta \sin \alpha}
{\displaystyle \log (m_{{\tilde t}_2}^2 / m_{{\tilde t}_1}^2)
 \over (m_{{\tilde t}_2}^2 - m_{{\tilde t}_1}^2)}   \ ,
\end{eqnarray}
where
\begin{equation}
	m^2_A = - {3 \lambda m_t^2 A_t \cos \phi_t  \over 8 \pi^2 v \sin 2 \alpha  \sin^2 \beta} f (m_{{\tilde t}_1}^2,  \ m_{{\tilde t}_2}^2)      \ ,
\end{equation}
\begin{eqnarray}
\Delta_{{\tilde t}_1} & = & \lambda x \cot \beta - A_t \cos \phi_t  \  , \cr
& & \cr
\Delta_{{\tilde t}_2} & = & \lambda x \cot \beta \cos \phi_t - A_t \ , \cr
& & \cr
\Delta_{\tilde t} & = & \left ( {4 m_W^2 \over 3} - {5 m_Z^2 \over 6} \right)
        \left \{(m_Q^2 - m_T^2) + \left ( {4 m_W^2 \over 3} - {5 m_Z^2 \over 6} \right) \cos 2 \beta \right \}  \ ,
\end{eqnarray}
\begin{eqnarray}
G_a & = & {g_1^{' 2} \over 36} (4 C_{2 \theta} - 1) - {g_1^{''2} \over 36} (4 C_{2 \theta} + 1) \ , \cr
& & \cr
G_b & = & {g_1^{' 2} \over 36} (\sqrt{15} S_{2 \theta} + C_{2 \theta} - 4)
- {g_1^{''2} \over 36} (\sqrt{15} S_{2 \theta} + C_{2 \theta} + 4)  \ , \cr
& & \cr
G_c & = &\mbox{} - {g_1^{' 2} \over 18} (\sqrt{15} C_{\theta} - 5 S_{\theta} ) S_{\theta}
+ {g_1^{''2} \over 18} (\sqrt{15} S_{\theta} + 5 C_{\theta} ) C_{\theta} \ , \cr
& & \cr
G_d & = & {g_1^{' 2} \over 72} (10 - 3 \sqrt{15} S_{2 \theta} + 5 C_{2 \theta})
+ {g_1^{''2} \over 72} (10 + 3 \sqrt{15} S_{2 \theta} - 5 C_{2 \theta})  \ ,
\end{eqnarray}
and the dimensionless function $g$ is defined as
\begin{equation}
 g(m_x^2,m_y^2) = {m_y^2 + m_x^2 \over m_x^2 - m_y^2} \log {m_y^2 \over m_x^2} + 2 \ .
\end{equation}

Note first that the matrix elements $M_{i5}^1$ ($i = 1,2,3,4$)
are proportional to $\sin \phi_t$.
If $\phi_t = 0$, these elements would be zero, and the scalar-psuedoscalar mixing
at the one-loop level would not occur in the Higgs sector.
Therefore, there would be no CP violation in the Higgs sector at the one-loop level.
The squared mass of the pseudoscalar Higgs boson at the one-loop level, $m_P^2$,
would be given simply by the $(5,5)$-th element of the $M$, taking $\phi_t =0$.
It is given by adding the radiative corrections as
\begin{equation}
	m_P^2 =  2 \lambda A v {\cos \phi  \over \sin 2 \alpha}
- {3 \lambda m_t^2 A_t  \over 8 \pi^2 v \sin 2 \alpha  \sin^2 \beta} f (m_{{\tilde t}_1}^2,  \ m_{{\tilde t}_2}^2)   \ .
\end{equation}
In this case, the $D$-terms of extra $U(1)$ symmetries
would not contribute to the mass of the pseudoscalar Higgs boson either
at the tree level or at the one-loop level.

If, on the other hand, $\phi_t \ne 0$, there would be CP violation at the one-loop level,
The CP phase in the radiative corrections generates the scalar-pseudoscalar mixing,
thus the five neutral Higgs bosons are no longer states of definite CP parity.
In this case, the mass matrix should be diagonalized
to obtain mass eigenstates $h_i$ ($i = 1,2,3,4,5$)
whose squared masses $m_{h_i}^2$ ($i = 1,2,3,4,5$) are the eigenvalues of the mass matrix.
These five neutral Higgs bosons are usually numbered
such that $h_1$ is the lightest neutral Higgs boson and $h_5$ is the heaviest.
Hereafter, we work in the explicit CP violation scenario, that is, with $\phi_t \ne 0$.

In our model, the squared masses of the two extra gauge bosons $m_{Z^{'}}^2$ and $m_{Z^{''}}^2$ are
obtained as the eigenvalues of the mass matrix for them.
The explicit expressions for $m_{Z^{'}}^2$ and $m_{Z^{''}}^2$ are given as
\begin{eqnarray}
m_{Z^{'}}^2 & = & {1 \over 2} (m_Z^2 + m_{Z_1}^2) + \sqrt{ m_{Z_1}^2- m_Z^2)^2 + 4 \Delta_1  } \ , \cr
m_{Z^{''}}^2 & = & {1 \over 2} (m_Z^2 + m_{Z_2}^2) + \sqrt{ m_{Z_2}^2- m_Z^2)^2 + 4 \Delta_2  } \ ,
\end{eqnarray}
where
\begin{eqnarray}
m_{Z_1}^2 & = & {1 \over 9} g_1^{' 2} v^2  ( 4 - C_{2 \theta} + \sqrt{15} \cos 2 \beta S_{2 \theta} )
+ {20 \over 9} g_1^{' 2} x_1^2 S_{\theta}^2  \cr
& &\mbox{} + {5 \over 36} g_1^{' 2} x_2^2 ( 8 + 7 C_{2 \theta} - \sqrt{15} S_{2 \theta} ) \ , \cr
m_{Z_2}^2 & = & {1 \over 9} g_1^{''2} v^2 ( 4 + C_{2 \theta} - \sqrt{15} \cos 2 \beta S_{2 \theta} )
+ {20 \over 9} g_1^{''2} x_1^2 C_{\theta}^2  \cr
& &\mbox{} + {5 \over 36} g_1^{''2} x_2^2 (C_{\theta} + \sqrt{15} S_{\theta} )^2   \ , \cr
\Delta_1 & = & {1 \over 3} g_1^{'}  m_Z v (\sqrt{5} \cos 2 \beta S_{\theta} + \sqrt{3} C_{\theta} )   \ , \cr
\Delta_2 & = & {1 \over 3} g_1^{''} m_Z v (\sqrt{5} \cos 2 \beta C_{\theta} - \sqrt{3} S_{\theta} )   \  .
\end{eqnarray}
The two mixing angles in our model, $\alpha_1$ between $Z$ and $Z'$ and $\alpha_2$
between $Z$ and $Z''$, are expressed as
\begin{equation}
\alpha_i  =  {1 \over 2} \tan^{-1} \left ( {2 \Delta_i \over m_{Z_i}^2 - m_Z^2 } \right ) \ ,
\end{equation}
 for $i = 1,2$.

\section{NUMERICAL ANALYSIS}

For the sake of simplicity, we take $g_1^{'} = g_1^{''} = \sqrt{5/3} g_1$ in our numerical analysis,
motivated by the gauge coupling unification.
We consider the region of the parameter space bounded as $0 < \theta < \pi/2$,
$0 < \phi_t < \pi$, $1 < \tan \beta \leq 30$, and $0 < \lambda \leq 0.83$.
We assume that the lighter stop quark is heavier than the top quark.
We also assume that all of the relevant mass parameters, $m_P$, $m_Q$, $m_T$,
and $A_t$, vary within the range of 100 to 1000 GeV.
Note that we employ the mass of the pseudoscalar Higgs boson at the one-loop level
in the CP conserving scenario, $m_P$, instead of $A$ as an input parameter.
Further, we use a combined constraint of $\lambda x_1 > 150$ GeV,
as the experimental data on the chargino system set the lower bound on the effective $\mu$ parameter,
$\mu \equiv \lambda x_1$.
For the values of $x_1$ and $x_2$, we would set their ranges not by hand but by experimental constraints.

There are strong experimental constraints on the mass of the extra neutral gauge boson and the mixing
between $Z$ in the SM and the extra neutral gauge boson.
Thus, any model with extra neutral gauge bosons, such as our model, should comply with these constraints,
whose exact values may dependent on the specific structures of the models.
We would like to take in this article that the mixing angles, $\alpha_1$ and $\alpha_2$,
should be smaller than $3 \times 10^{-3}$ and the masses of the two extra gauge bosons,
$m_{Z^{'}}^2$ and $m_{Z^{''}}^2$, should be larger than 800 GeV.

The experimental constraints on the Higgs sector should also be taken into account
in the numerical analysis.
The latest experimental analyses tell that the SM Higgs boson lighter than 114.5 GeV is
excluded at the 95 \% confidence level.
This lower bound on the SM Higgs boson mass may be applied to our model
by considering the relevant Higgs couplings.
Recently, the LEP collaborations reported the model-independent upper bound
on $(g_{ZZH}/g_{ZZH}^{\rm SM})^2$ at the 95 \% confidence level [16].

First, we determine the values of $x_1$ and $x_2$,
varying the values of other relevant parameters within their allowed ranges.
While the values of the above parameters are chosen by the random number generation method
within their respective ranges, the values of $x_1$ and $x_2$ are determined
in terms of the other parameters by imposing experimental constraints.
The result is shown in Fig.1(a), where a distribution of 7125 points is displayed
in the ($x_1$,$x_2$)-plane.
These points are selected among $10^5$ random points as they satisfy all of the above experimental constraints.
Each point represents a set of parameter values,
of which the values of $x_1$ and $x_2$ are explicit while others are implicit.
It is notable that these selected points are distributed in the area of the ($x_1$,$x_2$)-plane
where $x_1+x_2 \ge 2100$ GeV.
Some points are scattered at $x_1 \sim 400$ GeV.

Then, we calculate for each point in Fig.1(a) the mass of the lightest neutral Higgs boson
in our model.
In this way, it is clear that the results are consistent
with the relevant parameter ranges as well as the experimental constraints.
The result is shown in Fig.1(b).
It is quite remarkable that the majority of the points are scattered
within the range of $117 \le m_{h_1} \le 140$ GeV, while a few of them are distributed
where $m_{h_1}$ is as low as 30 GeV.
The result of Fig.1(b) suggests that the mass of the lightest neutral Higgs boson in our model
is most probably about 130 GeV at the one-loop level.

One may notice some pattern in Fig.1(b).
We find that this pattern comes from the experimental constraints on the extra gauge bosons
rather than that the experimental bound on the SM Higgs boson mass.
The lower bound on $x_1+x_2$ is found to arise from the experimental constraints
on the masses of the extra neutral gauge bosons.
Meanwhile, most of points with $m_{h_1} < 115$ GeV are excluded
by the experimental constraints on the SM Higgs boson mass.
We also calculate the masses of other neutral Higgs bosons.
The results are shown in Figs. 1(c) and (d), where we display the correlation
between $m_{h_3}$ and $m_{h_2}$ in Fig. 1(c) and
the correlation between $m_{h_5}$ and $m_{h_4}$ in Fig. 1(d).
The points in these figures are obtained with the same parameter values as in Figs. 1(a) or (b).
Note the clear hierarchy between the masses of the neutral Higgs bosons
such that $m_{h_3} > m_{h_2}$ in Fig. 1(c) and $m_{h_5} > m_{h_4}$ in Fig. 1(d).
The ranges for the masses of heavier neutral Higgs bosons in our model,
estimated using the aforementioned parameter values, are:
$100 < m_{h_2} < 997$ GeV, $116 < m_{h_3} < 998$ GeV, $262 < m_{h_4} < 1189$ GeV,
and $987 < m_{h_5} < 1536$ GeV, where the upper bounds come from theoretical arguments
and the lower bounds come from phenomenological constraints.

Now, we examine the possibility of discovering one of the neutral Higgs bosons
in our model in the $pp$ collisions at the LHC, where the most dominant process
for the Higgs production is the gluon fusion process, with thick QCD backgrounds.
The $WW$ fusion process is considered as the next dominant process for the Higgs production,
which is relatively cleaner than the gluon fusion process.
We would like to focus on the $WW$ fusion process.

We find that the PYTHIA program is useful for calculating the Higgs production mechanism than
for other processes, although it has not yet been applied to the CP violation scenario
in the MSSM Higgs sector.
However, the production cross section of the neutral Higgs bosons in our model
with explicit CP violation via the $WW$ fusion process in $pp$ collisions
is obtained by using the PYTHIA 6.4 program after appropriately
modifying the relevant Higgs coupling coefficients [17].
More precisely, we normalize $G_{WWH_i}$, the $WWh_i$ coupling coefficient of the Higgs coupling
to a pair of $W$ bosons, by the corresponding SM Higgs coupling coefficient.
We have
\begin{equation}
G_{WWh_i} = (\cos \beta O_{1i} + \sin \beta O_{2i}) \ ,
\end{equation}
where $O_{ij}$ ($i,j = 1,2,3,4,5$) are the elements of the orthogonal matrix
that diagonalizes the mass matrix for the five neutral Higgs bosons.

Technically, we set the number of events to generate for each point as NEV = 2000.
The Higgs coupling coefficient is set by MSTP(4)=1, and the normalized Higgs coupling
to a $W$ boson pair is set by PARU(165) = $G_{WWh_1}$.
The factorization scale and the renormalization scale are taken
to be the neutral Higgs boson mass, that is,
PARP(193) = PMAS(25,1) and PARP(194)= PMAS(25,1).
The PDF library of the CTEQ5L is used in our program, MSTP(51)=7,
which is the default parton distribution function set for the proton in PYTHIA 6.4.
We use MSTP(33)=0 to include the $K$ factor in hard cross sections for parton interactions in
PYTHIA 6.4 by default.
The $WW$ fusion process for the lightest neutral Higgs boson is set by MSUB(124) = 1.

In this way, we obtain all of $\sigma_{WWh_i}$ ($i = 1,2,3,4,5$), the production cross sections of $h_i$
in our model
with explicit CP violation via the $WW$ fusion process in $pp$ collisions.
They are given as functions of the participating neutral Higgs boson masses.
Among the five production cross sections, we select the largest one,
as we are interested in discovering any one of the five neutral Higgs bosons.
Thus, we introduce
\begin{equation}
\sigma_{WWh} = {\rm MAX} (\sigma_{WWh_1}, \sigma_{WWh_2}, \sigma_{WWh_3}, \sigma_{WWh_4}, \sigma_{WWh_5}) \ .
\end{equation}

We show our result in Fig. 2, where we plot $\sigma_{WWh}$ against $m_{h_1}$.
The parameter values for each point are the same as in Fig.1(a) or Fig.1(b).
We find that the smallest value for $\sigma_{WWh}$ is about 1 pb.
This implies that at least one of the five neutral Higgs bosons in our model may be produced
with its cross section larger than 1 pb.
The accumulated integrated luminosity of $30^{-1}$ fb at the LHC would yield
6000 raw Higgs events, if we allow  20 \% for the efficiency and acceptance.
Therefore, we expect with relatively strong confidence that at least one of the five neutral Higgs bosons
in our model might be produced via the $WW$ fusion process at the LHC, if they exist.

Here, the roles that the exotic quarks take part in are worth mentioning
with respect to the Higgs phenomenology of our model.
The exotic quarks may inhabit the fundamental 27 representation of $E_6$,
which is the underlying gauge symmetry of our model.
In the fundamental 27 representation, 15 components are occupied by the SM matter fields,
4 components by the two Higgs doublets, 2 components by the Higgs singlet,
and the remaining 6 components are occupied by the exotic quarks [12-14,18-20].

The form of the superpotential tells us that the exotic quarks may couple to various Higgs fields.
They couple directly to the neutral component of the Higgs singlet $N_1$
and indirectly, through the mixing among the neutral Higgs bosons via the diagonalization matrix,
to other neutral Higgs fields.
If the masses of the exotic quarks are comparable to the electroweak symmetry breaking or
SUSY breaking scales, the low energy SUSY phenomenology might be affected by their presence.
The effects of the exotic quarks might appear in the gluon fusion processes for Higgs productions,
as well as in the Higgs decay processes.
In particular, for example, the Higgs decays into a pair of gluons or photons might
receive the effects of the exotic quarks, when the mass of the Higgs boson is
below the range where the decay channel into a pair of gauge bosons are not yet open.

However, it is somewhat difficult to predict the amount of the exotic quark effects
because it depends on the relevant parameters in a complicated way.
The coupling strength of the exotic quarks to the neutral Higgs bosons are weak
in general but might be strong, depending on what is the explicit structure of the orthogonal matrix
that diagonalizes the mass matrix for the neutral Higgs bosons.
Therefore, it would be valuable to study elsewhere a comprehensive research
on the effects of the exotic quarks in our model.

\section{CONCLUSIONS}

We study a supersymmetric $E_6$ model with two extra $U(1)$ symmetries besides the SM gauge symmetry,
and two neutral Higgs singlets besides two MSSM Higgs doublets.
We find that the Higgs sector of our model may generally accommodate
a non-trivial complex phase which can cause the scalar-pseudoscalar mixing
among the five neutral Higgs bosons, by virtue of radiative corrections
due to the top and stop quark loops.
Thus, explicit CP violation at the one-loop level is viable in our model.

Numerical analysis shows that there are parameter regions in our model which comply
with a number of experimental constraints such as the lower bound
on the extra neutral gauge boson masses
and the upper bound on the mixing between the extra neutral gauge bosons and the SM neutral gauge boson.
Within the allowed parameter regions,
we study the behavior of the vacuum expectation values of the two Higgs singlets, $x_1$ and $x_2$.
We find that they cannot be simultaneously small.
The experimental constraints on the extra neutral gauge bosons restrict
that $x_1+x_2$ should be larger than 2100 GeV
whereas either one of them may be as small as 400 GeV.

The possibility of discovering one of the five neutral Higgs bosons in our model is
examined by calculating the production cross sections using the PYTHIA 6.4 program,
where the relevant Higgs couplings are modified suitably.
We focus the $WW$ fusion process at the LHC for their productions.
We find that at least one of five neutral Higgs bosons can be produced enough
via the $WW$ fusion process at the LHC.
Thus, we speculate that the present SUSY $E_6$ model can be tested by the Higgs searches at the LHC.

\vskip 0.3 in
\noindent
{\large {\bf ACKNOWLEDGMENTS}}
\vskip 0.2 in

S. W. Ham thanks his late teacher, Bjong Ro Kim, for learning supersymmetry.
He is supported by the Korea Research Foundation Grant funded
by the Korean Government (MOEHRD, Basic Research Promotion Fund) (KRF-2007-341-C00010).
He was supported by grant No. KSC-2008-S01-0011 from Korea Institute of Science and Technology Information.
This work is supported by Konkuk University in 2007.


\newpage

\vfil\eject


{\large {\bf FIGURE CAPTION}}
\vskip 0.3 in
\noindent
FIG. 1(a). : A distribution of 7125 points in the $(x_1,x_2)$ plane.
Each point represents a set of parameter values that satisfies the experimental constraints
on the extra neutral gauge boson masses, on their mixings with the SM neutral gauge bosson,
and on the SM Higgs boson mass.
The values of $x_1$ and $x_2$ are explicitly shown, and the other parameters have certain values
within their ranges respectively by the random number generation method:
$1 < \tan \beta \leq 30$, $0 < \lambda \leq 0.83$, $0  < \theta < \pi/2$, $0 < \phi_t < \pi$,
$100 \leq m_A, m_Q, m_T, A_t \leq 1000$ GeV.

\vskip 0.3 in
\noindent
FIG. 1(b). : The plot of $m_{h_1}$ against $(x_1 + x_2)$.
For each of the 7125 points in Fig. 1(a), the mass of the lightest neutral Higgs boson is
calculated in terms of the parameter values represented by the point.

\vskip 0.3 in
\noindent
FIG. 1(c). : The distribution of 7125 points in the $(m_{h_3}, m_{h_2})$-plane.
They are distributed between $100 < m_{h_2} < 997$ GeV and $116 < m_{h_3} < 998$ GeV,
and they satisfy $m_{h_3} > m_{h_2}$.
These points are obtained with the same parameter values as in Figs. 1(a) or (b).

\vskip 0.3 in
\noindent
FIG. 1(d). : The distribution of 7125 points in the $(m_{h_5}, m_{h_4})$-plane.
They are distributed between $262 < m_{h_4} < 1189$ GeV and $987 < m_{h_5} < 1536$ GeV,
and they satisfy $m_{h_5} > m_{h_4}$.
These points are obtained with the same parameter values as in Figs. 1(a) or (b).

\vskip 0.3 in
\noindent
FIG. 2. : The polt of $\sigma_{WWh}$ against $m_{h_1}$.
The production cross sections of the five neutral Higgs bosons via $WW$ fusion process
in $pp$ collisions are calculated in terms of the parameter values represented by the point,
and the largest of them is chosen, for each of the 7125 points in Fig. 1(a).

\vfil\eject

\setcounter{figure}{0}
\def\figurename{}{}%
\renewcommand\thefigure{FIG. 1(a)}
\begin{figure}[t]
\begin{center}
\includegraphics[scale=0.6]{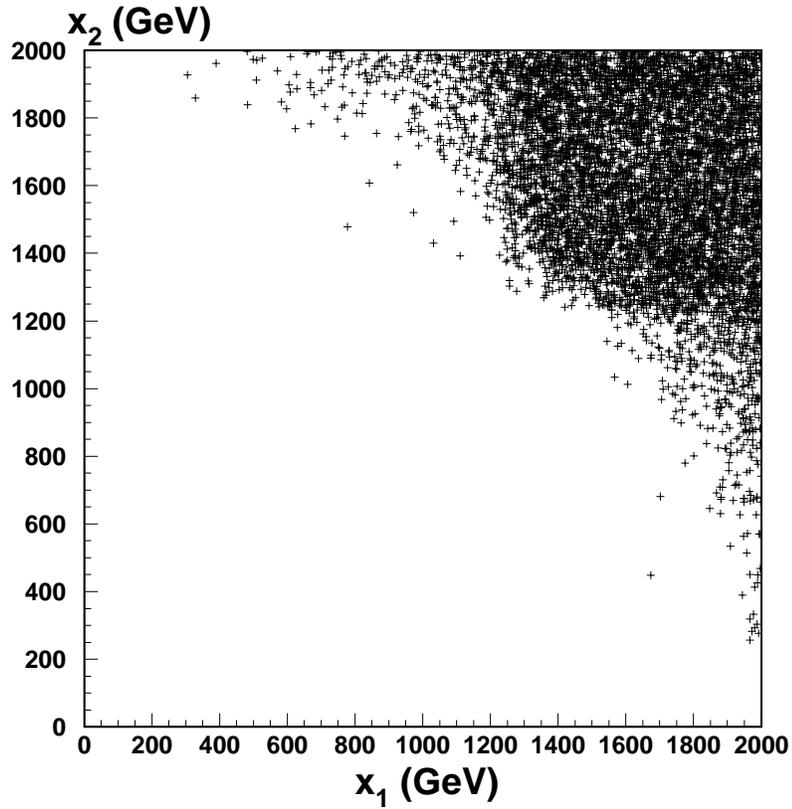}
\caption[plot]{A distribution of 7125 points in the $(x_1,x_2)$ plane.
Each point represents a set of parameter values that satisfies the experimental constraints
on the extra neutral gauge boson masses, on their mixings with the SM neutral gauge bosson,
and on the SM Higgs boson mass.
The values of $x_1$ and $x_2$ are explicitly shown, and the other parameters have
certain values within their ranges respectively by the random number generation method:
$1 < \tan \beta \leq 30$, $0 < \lambda \leq 0.83$, $0  < \theta < \pi/2$, $0 < \phi_t < \pi$,
$100 \leq m_A, m_Q, m_T, A_t \leq 1000$ GeV.
}
\end{center}
\end{figure}

\renewcommand\thefigure{FIG. 1(b)}
\begin{figure}[t]
\begin{center}
\includegraphics[scale=0.6]{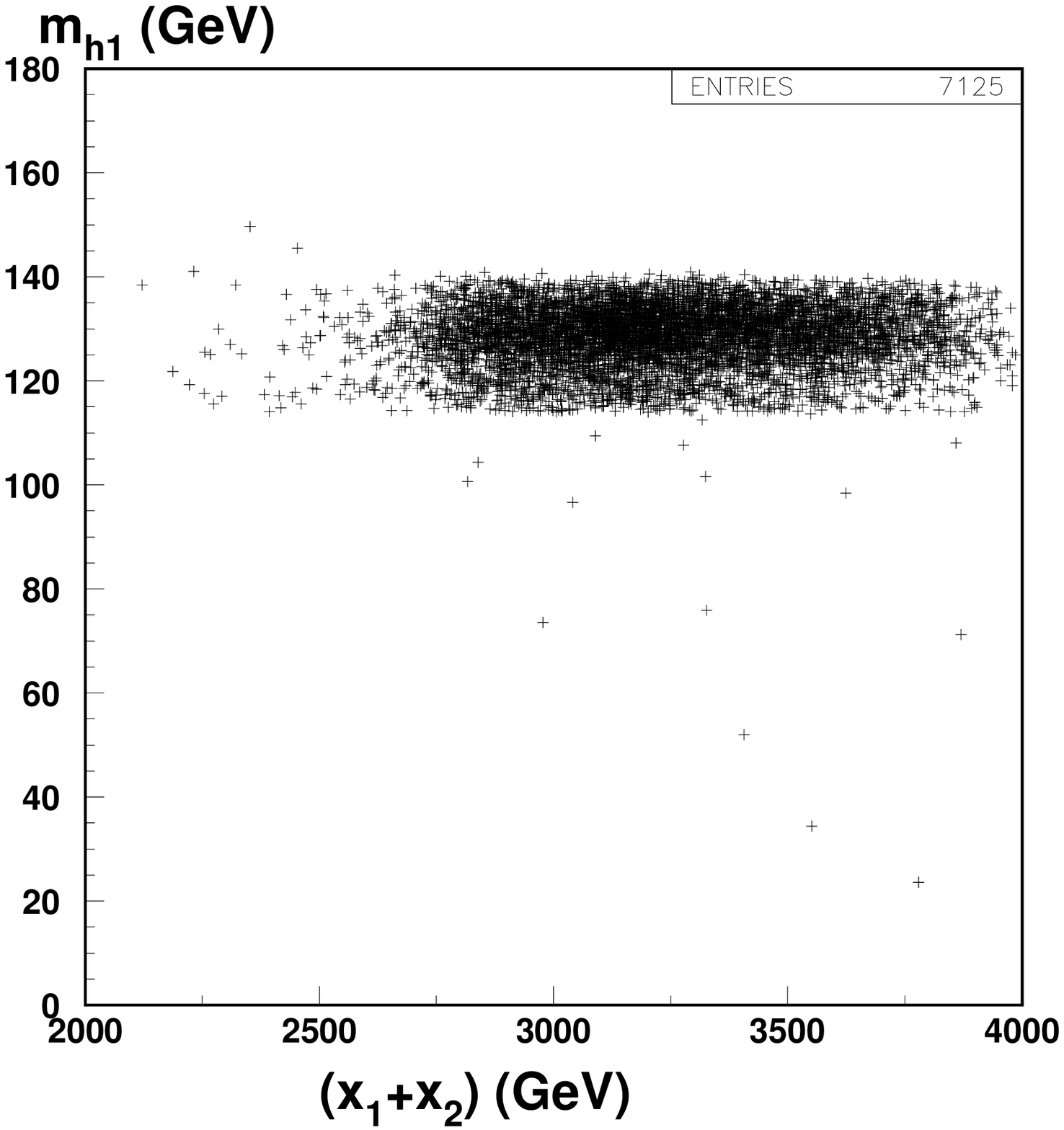}
\caption[plot]{The plot of $m_{h_1}$ against $(x_1 + x_2)$.
For each of the 7125 points in Fig. 1(a),
the mass of the lightest neutral Higgs boson is calculated in terms of the parameter values
represented by the point. }
\end{center}
\end{figure}

\renewcommand\thefigure{FIG. 1(c)}
\begin{figure}[t]
\begin{center}
\includegraphics[scale=0.6]{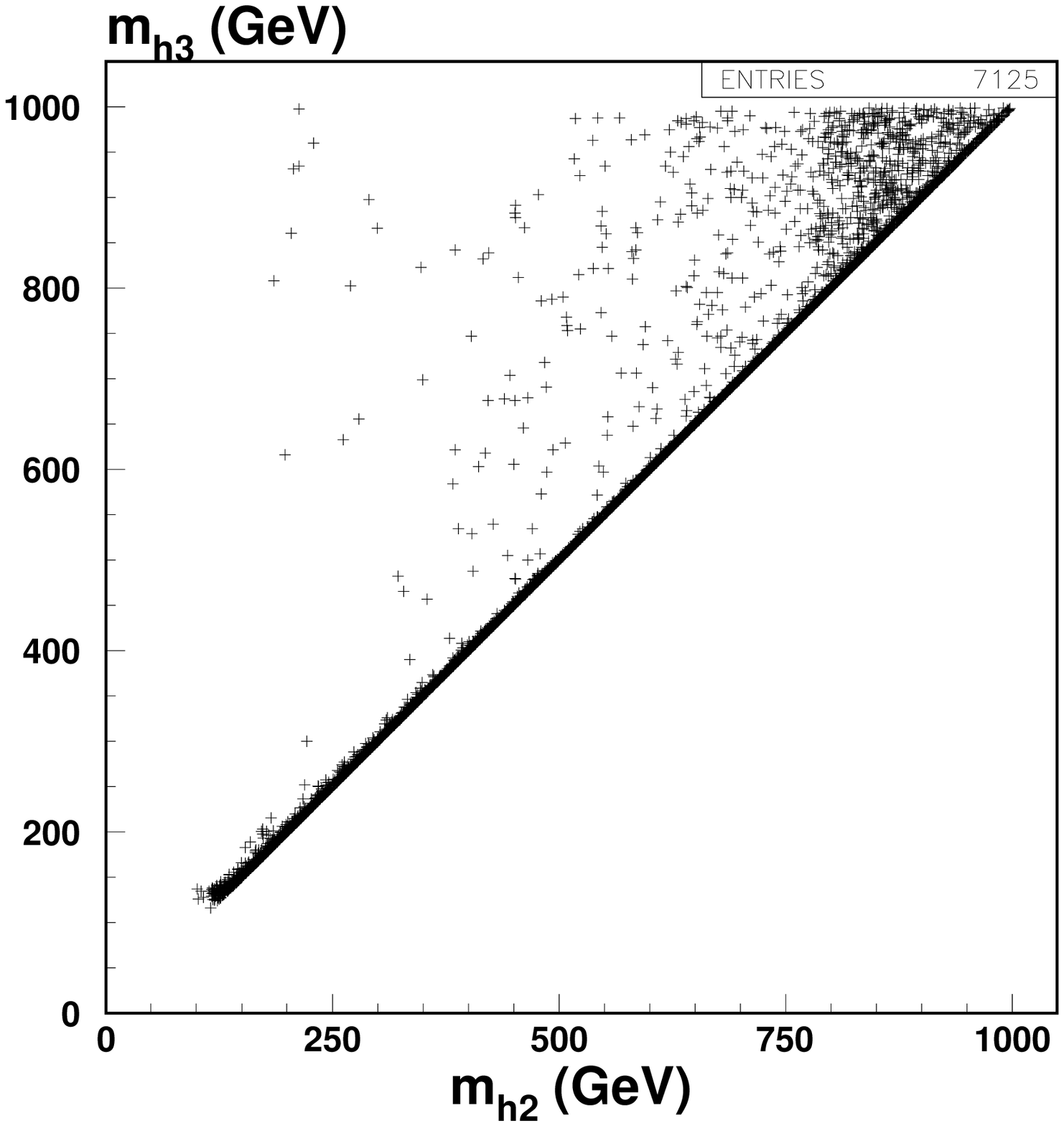}
\caption[plot]{The distribution of 7125 points in the $(m_{h_3}, m_{h_2})$-plane.
They are distributed between $100 < m_{h_2} < 997$ GeV and $116 < m_{h_3} < 998$ GeV,
and they satisfy $m_{h_3} > m_{h_2}$.
These points are obtained with the same parameter values as in Figs. 1(a) or (b).}
\end{center}
\end{figure}

\renewcommand\thefigure{FIG. 1(d)}
\begin{figure}[t]
\begin{center}
\includegraphics[scale=0.6]{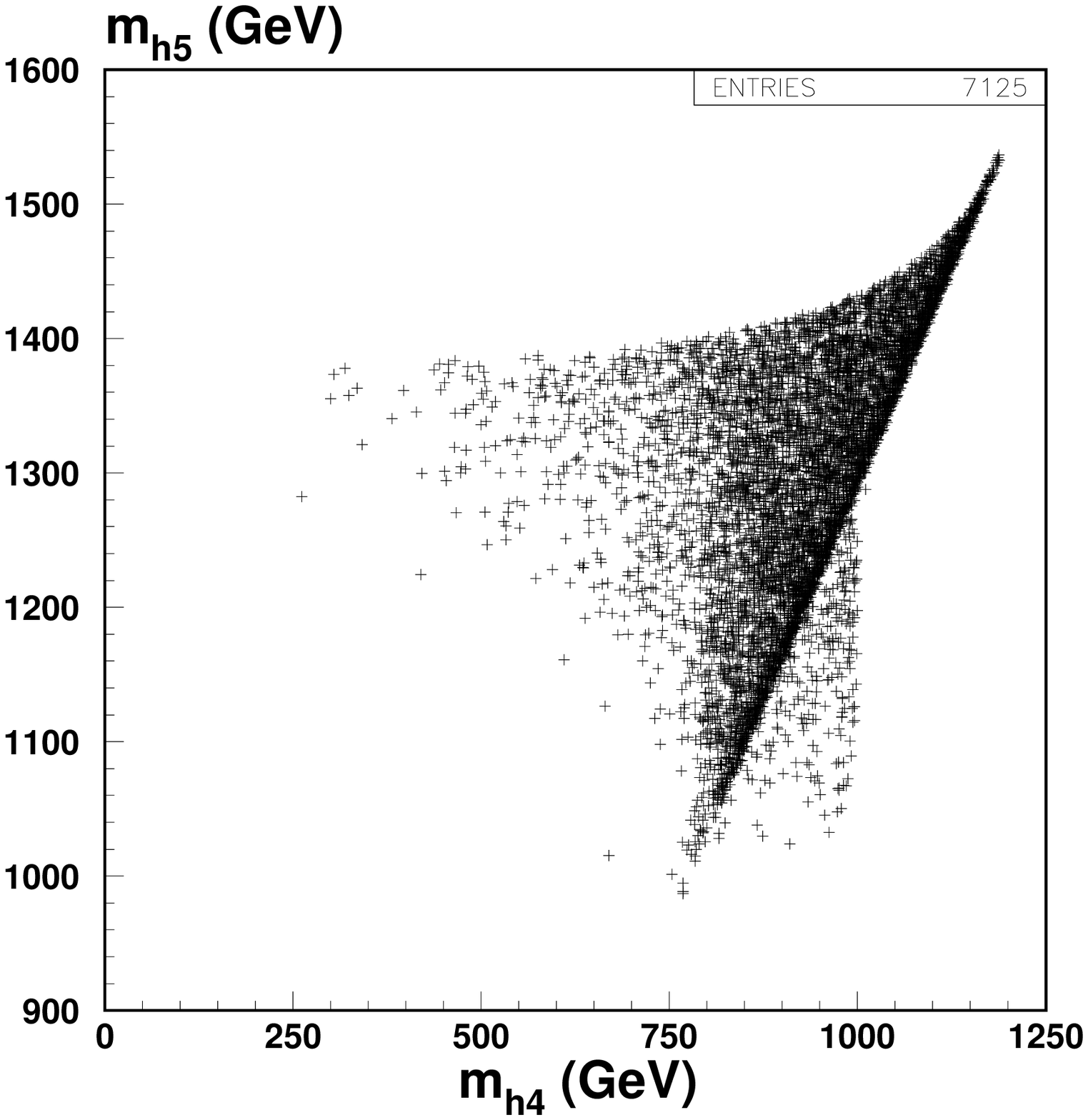}
\caption[plot]{The distribution of 7125 points in the $(m_{h_5}, m_{h_4})$-plane.
They are distributed between $262 < m_{h_4} < 1189$ GeV and $987 < m_{h_5} < 1536$ GeV,
and they satisfy $m_{h_5} > m_{h_4}$.
These points are obtained with the same parameter values as in Figs. 1(a) or (b).}
\end{center}
\end{figure}

\renewcommand\thefigure{FIG. 2}
\begin{figure}[t]
\begin{center}
\includegraphics[scale=0.6]{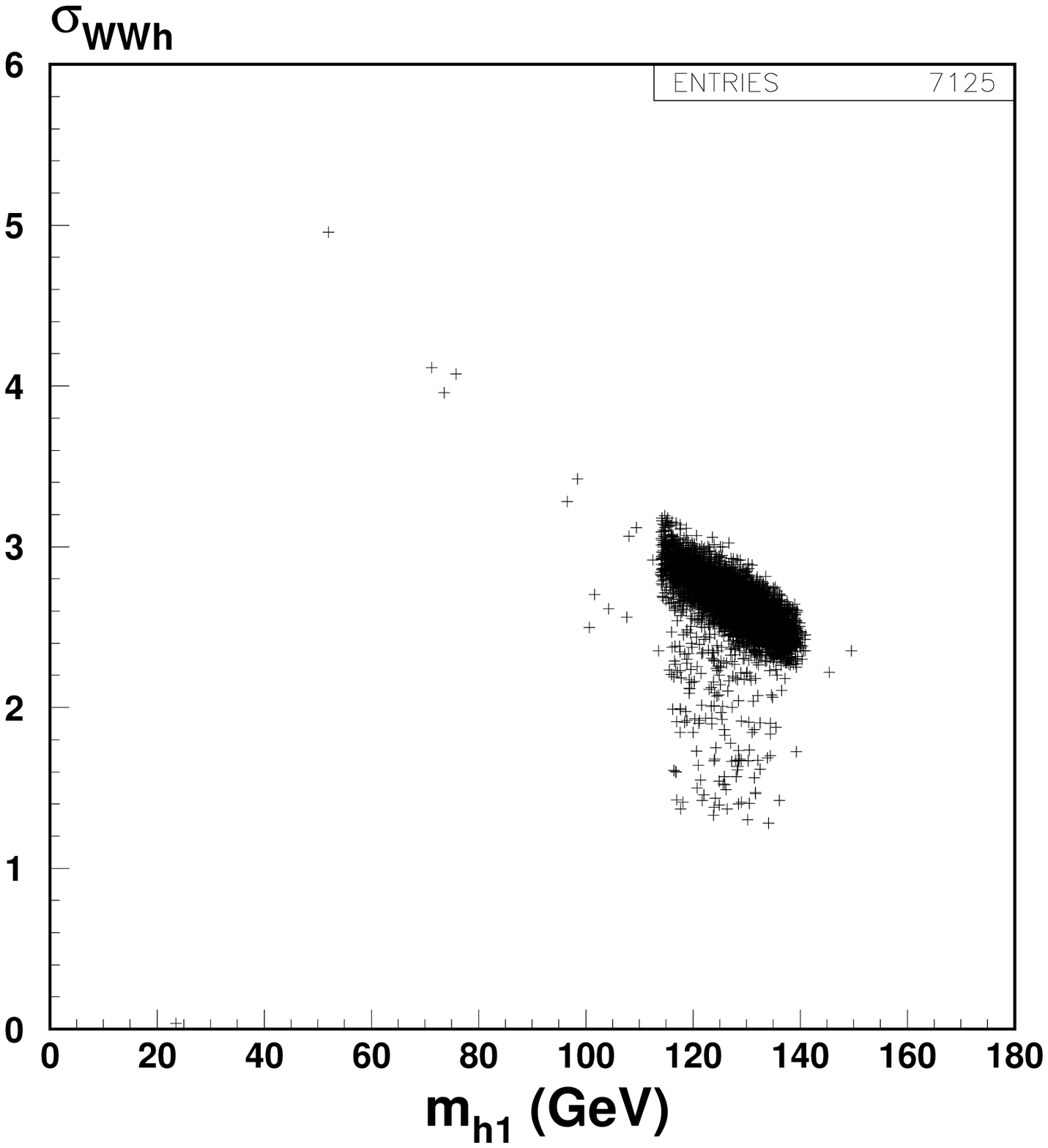}
\caption[plot] {The polt of $\sigma_{WWh}$ against $m_{h_1}$.
The production cross sections of the five neutral Higgs bosons via $WW$ fusion process
in $pp$ collisions are calculated in terms of the parameter values represented by the point,
and the largest of them is chosen, for each of the 7125 points in Fig. 1(a). }
\end{center}
\end{figure}

\end{document}